\documentclass[aps,twocolumn,prl,floatfix]{revtex4}

\usepackage{amsmath}
\usepackage{epsfig}
\usepackage{graphicx}

\newcommand{\beq}{\begin{equation}}
\newcommand{\eeq}{\end{equation}}
\newcommand{\bea}{\begin{eqnarray}}
\newcommand{\eea}{\end{eqnarray}}

\def\leq{\raise 0.4ex\hbox{$<$}\kern -0.8em\lower 0.62ex\hbox{$-$}}
\def\geq{\raise 0.4ex\hbox{$>$}\kern -0.7em\lower 0.62ex\hbox{$-$}}
\def\lsim{\raise 0.4ex\hbox{$<$}\kern -0.8em\lower 0.62ex\hbox{$\sim$}}
\def\gsim{\raise 0.4ex\hbox{$>$}\kern -0.7em\lower 0.62ex\hbox{$\sim$}}
\def\pm{\,\raise 0.4ex\hbox{$+$}\kern -0.8em\lower 0.62ex\hbox{$-$}\,}

\begin{document}

\title{Reply to ``Comment on `Black Holes are neither Particle Accelerators nor Dark Matter Probes' ''}


\maketitle

A 2009 result in the study of geodesics in Kerr spacetime has stirred a debate, primarily centered on the practical relevance of the result, hereafter called the ``BSW effect'' \cite{Ban1}.
The effect, which can be summarized as the divergence of the center-of-mass (CM) energy for colliding particles as they approach the Kerr horizon, has been suggested as a means
of studying possible dark matter annihilations, or even probing the Planck scale.  
However, we showed that for the specific example explored in
a followup paper from the original BSW authors \cite{Ban2}, that
the energy reaching a distant observer in that case is actually a decreasing function of the collisional radius, and is therefore vanishingly small at radii where the BSW effect
predicts large CM energies.  This conclusion is correct; the photon formed from a collision just outside
the horizon can suffer a diverging redshift with decreasing radius at a rate that exceeds the divergence of the CM energy, and thereby results in a vanishing energy reaching infinity.
As stated in \cite{zas}, the outgoing energy of the escaping particle is given by
\beq
E_{\infty,\,{\rm tot}} = (\alpha \gamma + \ell \omega + u_r v_r) m_{\chi} \,,
\label{eq:zas} 
\eeq
where two particles with rest mass $m_{\chi}$ collide, with one particle having angular momentum and radial velocity of $\ell$ and $u_r$ and the other having angular and radial velocities
of $\omega$ and $v_r$ relative to local zero angular momentum observers (ZAMOs),
$\gamma$ is the relative Lorentz factor between the two colliding particles, and $\alpha \equiv d\tau /dt$ in this context is the gravitational redshift of the CM frame.
Only the first term in Eq.~\eqref{eq:zas} can be considered as energy resulting from the collision, and therefore from the BSW effect, since
it stems from the diverging term $2 \gamma m^2$ appearing in $E_{\rm cm}^2$ (see Eq.~3 below).  Nonetheless, Zaslavskii is quite right
that the true total energy at infinity must include the other terms.  Indeed, although our Eqs.~6 and 7 in \cite{STM} 
apply only to the BSW contribution to the total energy, we include the standard collisionless
contribution to the energy due only to the dynamics of any particle within the ergosphere of a Kerr black hole when we say that the maximum total energy reaching infinity as $r\rightarrow M$
is simply $m_{\chi}$, despite the fact that our Eq.~7 for $E_{\infty}$ clearly vanishes in this limit.  In this case, the total energy at infinity 
results solely from the term $m_{\chi} \ell \omega$, with max$(\ell)=2$ and $\omega=\Omega_{\rm H}$, where $\Omega_{\rm H}=1/2$ is the angular velocity of the horizon.
Unfortunately, in \cite{STM} we referred to both the true total energy at infinity, $E_{\infty,\,{\rm tot}}$, and the contribution from BSW, $E_{\infty,\,{\rm BSW}}$, by the same name
``$E_{\infty}$'', which no doubt has contributed to the confusion on this point.
In the case where $\ell=0$, for example, $E_{\infty,\,{\rm tot}} \propto r-1$, whereas $E_{\infty,\,{\rm BSW}} \propto (r-1)^{3/2}$ as given in Eq. 7 of \cite{STM}, indicating that
the BSW effect is the subdominant contribution to the total energy in this case.
The first term in Eq.~\ref{eq:zas} must be dominant and $E_{\infty} \approx \alpha \gamma m_{\chi}$ must be satisfied for the BSW effect to
have any hope of ``probing the Planck scale'' by yielding $E_{\infty,\,{\rm tot}} \gg m_{\chi}$.  As has been shown in \cite{Bejger}, however, this condition is never satisfied, 
and $0 \, \leq \,  E_{\infty,\,{\rm tot}} \, \lsim \, 2m_{\chi}$, so BSW can never probe energies larger than the rest mass
of the infalling particles. 

The second point of contention in \cite{zas} relates to confusion between the CM energy, $E_{\rm cm}$, and the energy observed by the local zero angular momentum (ZAMO) observers,
$E_{\rm loc}$.  Zaslavskii assumes that the collision products are massive particles, and therefore one must consider the rest masses of the individual particles, as well as
the relative Lorentz factor of the collision products when relating 
$E_{\rm loc}$ to $E_{\rm cm}$.  However, in \cite{STM} we explicitly assume that the collision creates two photons.  Therefore, we immediately find that 
\beq
E_{\rm cm}^2 = 2 k_{3\,\mu} k_4^{\,\, \mu} = E_{\rm loc}^2\,,
\eeq
is satisfied, contrary to the claim in \cite{zas},
where $k_{3\,\mu}$ and $k_{4\,\mu}$ are the wave four-vectors of photons 3 and 4, which satisfy $k_{3\,0} = k_{4\,0}$ and $k_{3\,i} = -k_{4\,i}$ for the case considered in \cite{STM,Ban2}
where the CM frame has zero angular momentum.  Furthermore, even for collision products with masses $m_3=m_4=m_{\chi}$, if we use Eq.~2 from \cite{zas}, and take the limit where the BSW effect dominates
the CM energy, we find
\bea
E_{\rm cm}^2 &=& m_3^2 + m_4^2 + 2m_3m_4\gamma(3,4) \nonumber \\
&\approx& 2m_3m_4\gamma(3,4)=\left(2m_{\chi}\gamma(3,{\rm CM})\right)^2 = E_{\rm loc}^2\,,
\eea
Therefore, our asymptotic behavior in Eq. 7 of \cite{STM} is correct.

\begin{figure}
\includegraphics[trim = 0mm 0mm 0mm 0mm, clip, width=.45\textwidth, angle=0]{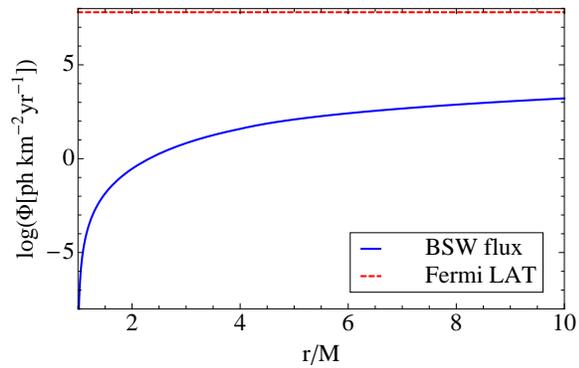}
\caption
{
Integrated flux $\Phi$ reaching an observer at  $D_L=10$ kpc from inside radius $r$ (solid line), compared to the flux sensitivity of the \emph{Fermi} LAT for a one year exposure (dashed line).
}
\label{fig:flux}
\end{figure}

To eliminate any possibility for further confusion regarding energies, observer frames, or particle types, we will focus here exclusively on the flux, which, as can easily be shown using the formalism in \cite{Ban2} with the key corrections
given in \cite{STM}, is many, many orders of magnitude too small to be observed with current (or foreseeable) instruments.  This was, in fact, the main conclusion we intended the reader to draw
from \cite{STM}, with the potentially vanishing energy of the miniscule number of escaping particles being, at most, a side note, particularly since variations on this point had already been made
by other authors in the BSW context.  In Fig.~\ref{fig:flux}, 
we show the integrated flux $\Phi$ reaching an observer at a distance of 10 kpc from inside radius $r$ (solid line).  This flux will include photons over the full range of allowed energies.
For comparison, we also show the flux sensitivity of the \emph{Fermi} Large Area Telescope (LAT) for a one year exposure, which is $\sim 2 \times 10^{-10}$ photons/cm$^2$/s
$= 6 \times 10^7$ photons/km$^2$/yr \cite{lat} (dashed line).  The flux from the region $r\, \leq\, 1.1\,M$, which is frequently used as a reference interval wherein the BSW effect is important, is
$\sim 16$ orders of magnitude too faint to be observed.  To put this in perspective, the LAT, with a collecting area of $\sim2\,{\rm m}^2$, would need to operate for $\sim10^{14}$ years to detect
a single BSW photon from a galactic source.

We hope to have clarified the points of apparent disagreement with \cite{STM} that were recently brought up by Zaslavskii in \cite{zas}.  Furthermore, we note 
our mutual agreement with Zaslavskii that the BSW effect yields no
observable flux.  Speaking for ourselves, and not necessarily for Zaslavskii, we feel that the lack of observable flux relegates this entire discussion to the realm of philosophy.  
Since this effect can never be observed in nature, it is therefore not a viable means
for studying dark matter annihilations, Planck scale physics, nor any other novel physical phenomena.

\end{document}